\documentclass[twocolumn,aps,prl,reprint,groupedaddress]{revtex4}
\usepackage{graphicx}
\usepackage{color}

\newcommand{\be}{\begin{equation}}
\newcommand{\ee}{\end{equation}}

\newcommand{\bea}{\begin{eqnarray}}
\newcommand{\eea}{\end{eqnarray}}

\newcommand{\la}{\langle}
\newcommand{\ra}{\rangle}

\newcommand{\lp}{\left(}
\newcommand{\rp}{\right)}

\renewcommand{\Im}{{\rm \, Im\,}}
\renewcommand{\vec}[1]{{\bf #1}}

\newcommand{\addJS}[1]{\textcolor{red}{#1}}
\newcommand{\addLL}[1]{\textcolor{blue}{#1}}

\begin{document}
\title{Photo-excited Carrier Dynamics and Impact Excitation Cascade in Graphene}

\author{Justin C. W. Song$^{1,2}$}
\author{Klaas J. Tielrooij$^3$}
\author{Frank H. L. Koppens$^3$}
\author{Leonid S. Levitov$^1$} 

\affiliation{$^1$ Department of Physics, Massachusetts Institute of Technology, Cambridge, Massachusetts 02139, USA}
\affiliation{$^2$ School of Engineering and Applied Sciences, Harvard University, Cambridge, Massachusetts 02138, USA}
\affiliation{$^3$ ICFO-Institut de Ci\'encies Fot\'oniques, Mediterranean Technology Park, Castelldefels (Barcelona) 08860 , Spain}





\begin{abstract}
Photo-excitation in solids can trigger a cascade in which multiple particle-hole excitations are generated.
We analyze the carrier multiplication cascade of impact excitation processes  in graphene and show that the number of pair excitations has a strong dependence on doping, which makes carrier multiplication gate-tunable. We also predict that the number of excited pairs as well as the characteristic time of the cascade scale linearly with photo-excitation energy. These dependences, as well as sharply peaked angular distribution of pair excitations, provide clear experimental signatures of carrier multiplication.
\end{abstract} 


\maketitle

Converting light to electrical currents or voltages is  a complex, multi-step process which involves photo-excited particle-hole pairs undergoing scattering by ambient charge carriers, by other photoexcited carriers and by lattice vibrations. One of the key questions in the field of optoelectronics is identifying materials in which {\it carrier multiplication} can occur, i.e. a single absorbed photon yielding a large number of particle-hole pairs as a result of the primary photoexcited pair producing secondary pairs. 
Efficient carrier multiplication relies on 
a combination of characteristics
such as a wide band of states with a large phase space density for pair excitations,  strong electron-electron scattering, and not too strong electron-phonon interaction.  While graphene is by no means a unique example of a system with these properties,
it is believed to fit the bill better than other materials.
This has motivated an intense investigation of photoexcitation processes in graphene-based systems \cite{kampfrath05,winzer10,winzer12,winnerl11,george08,lui10, rana, aleiner, butscher07, vasko08, kim11, breusing09}.

One aspect of graphene that distinguishes it from other materials is its truly two-dimensional structure which renders electronic states fully exposed. Photoexcitation in such a system generates photoexcited carriers that can in principle be extracted by a vertical transfer process, e.g. in a sandwich-type  tunneling structure. Vertical carrier extraction eliminates carrier loss in a lateral transport betwen photoexcitation region and contacts, often an important limiting factor for optoelectronic response in semiconductor systems.

Despite intense interest, the photo-excitation cascade in graphene remains poorly understood. Theory predicts that the linear dispersion of charge carriers acquires a negative curvature due to electron-electron interactions, $d^2\epsilon(k)/dk^2<0$ \cite{Guinea1994}, which inhibits decay via electron-electron scattering in undoped graphene \cite{aleiner}.
However, while the prediction of negative curvature appears to be in agreement with transport measurements \cite{Geim2011}, ARPES experiments support the notion of interaction-mediated decay \cite{Lanzara2006, Rotenberg2007}; interaction-induced quasiparticle decay remains the subject of ongoing debate 
\cite{aleiner,rana, polini_privatecommunication}.


\begin{figure}[t]
\includegraphics[scale=0.33]{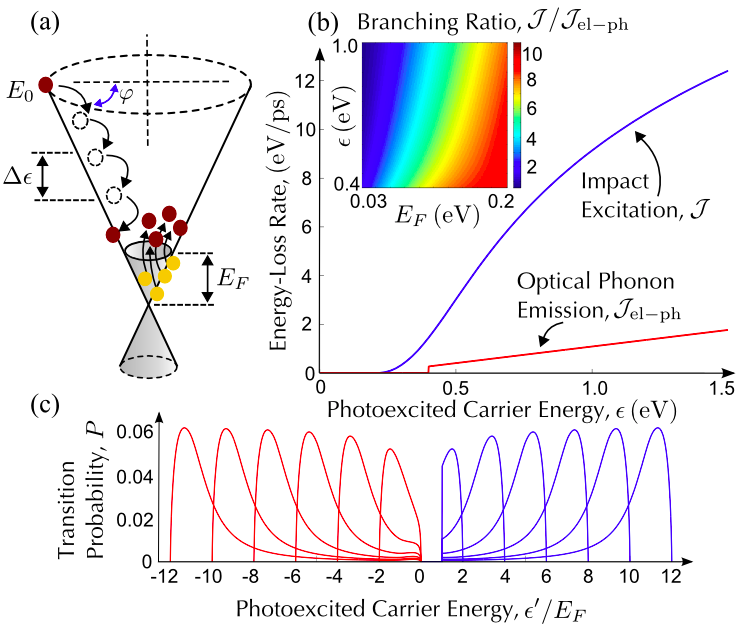}
\caption{ (a) Impact excitation (IE) cascade of a photo-excited carrier with initial energy $E_0$. Each cascade step involves electron-hole pair excitations with energy $\Delta \epsilon \sim E_F$, where $E_F$ is Fermi energy. The net number of  generated pairs
and the relaxation rate
depend strongly on $E_F$ [see Eqs.(\ref{eq:avN}),(\ref{eq:Jnumerical})], and thus can
be tuned by gate voltage. 
(b) Energy-loss rate 
via impact excitation, $\mathcal{J}$ [blue curve, see Eq.(\ref{eq:Jnumerical})],
and optical phonon emission, $\mathcal{J}_{\rm el-ph}$ \cite{appendix} (red curve) 
for a typical doping of $E_F = 0.2 \, {\rm eV}$. (Inset) Branching ratio, $\mathcal{J}/ \mathcal{J}_{\rm el-ph}$ vs. $\epsilon$ and $E_F$. 
 (c) Transition probability (in units of $\hbar^{-1}$) obtained from 
Eq.(\ref{eq:pav}).  IE processes with different initial energies, $\epsilon_i/E_F = -12, -10 , \dots, 10, 12$ (here $\Delta \epsilon = \epsilon_i - \epsilon'$). Electron (hole) contributions shown by blue (red) curves.}
\label{fig1}
\end{figure}

Besides its immediate utility for optolectronics, photo-excitation cascade in graphene is of interest because of its analogy with jets in particle physics, which are
narrow cones of hardrons and other particles produced in high-energy particle detectors. Massless dispersion in graphene limits the kinematics of photoexcited carriers undergoing scattering by other carriers
in much the same way as the  kinematics of hadronization of quarks or gluons in a particle physics experiment, where jets are created.
Because both the primary photoexcited carriers and the secondary carriers 
share the same linear dispersion, collinear
scattering is enhanced. As we will show, this results in sharp collimated angular features
reminiscent of ultra-relativisitic jets in high-energy physics.

Here we show that the photo-excitation cascade in doped graphene is distinct from that in undoped graphene. We identify {\it impact excitation} (IE) as the scattering process that dominates carrier relaxation dynamics in this system. Multiple secondary electron-hole (e-h) pairs produced by IE scattering involving a photo-excited carrier and ambient carriers in the Fermi-sea can lead to efficient carrier multiplication. Our analysis predicts that IE 
processes result in a  chain-like 
cascade consisting of sequential steps with relatively {\it small energy loss} per step $\Delta \epsilon \sim E_F$, where $E_F$ is the Fermi energy in graphene doped away from neutrality (see Fig.\ref{fig1}(a) and (c)). As we shall see, both the number of pairs produced in the cascade (carrier multiplication factor) and the characteristic energy for the pairs are highly sensitive to doping. As a result, the key parameters of photo-excitation cascade in graphene are expected to be gate tunable in a wide range.

As we shall see, the IE rate 
 takes highest values allowed by unitarity, $\Gamma\sim E_F/(2\pi\hbar)$. 
This fast characteristic rate makes this scattering process
a highly efficient relaxation pathway which dominates over the phonon-mediated pathway in a wide range of energies (see Fig.\ref{fig1}b).
Indeed, typical dopings of $E_F=200\, {\rm meV}$ yield $E_F/\hbar = (2\pi\times ) 0.48 \times 10^{14} \, {\rm Hz}$ which corresponds to a scattering time of $\tau \approx 20 \, {\rm fs}$ (this is consistent with the values 
found for the transport mean free path due to carrier-carrier scattering extrapolated to high energies \cite{tse08}).
This is far faster than electron-optical phonon scattering rates \cite{tse08,appendix}. 
As a result, IE produces an energy relaxation rate,
$\mathcal{J}$, that dominates over the energy relaxation rate from the emission of optical phonons, $\mathcal{J}_{\rm el-ph}$ (see Fig. \ref{fig1}(b)) for typical dopings and photo-excited carrier energies (see also Fig. \ref{fig1}(b) inset); IE controls the photo-excited carrier relaxation dynamics in graphene.

\begin{figure}[t]
\includegraphics[scale=0.18]{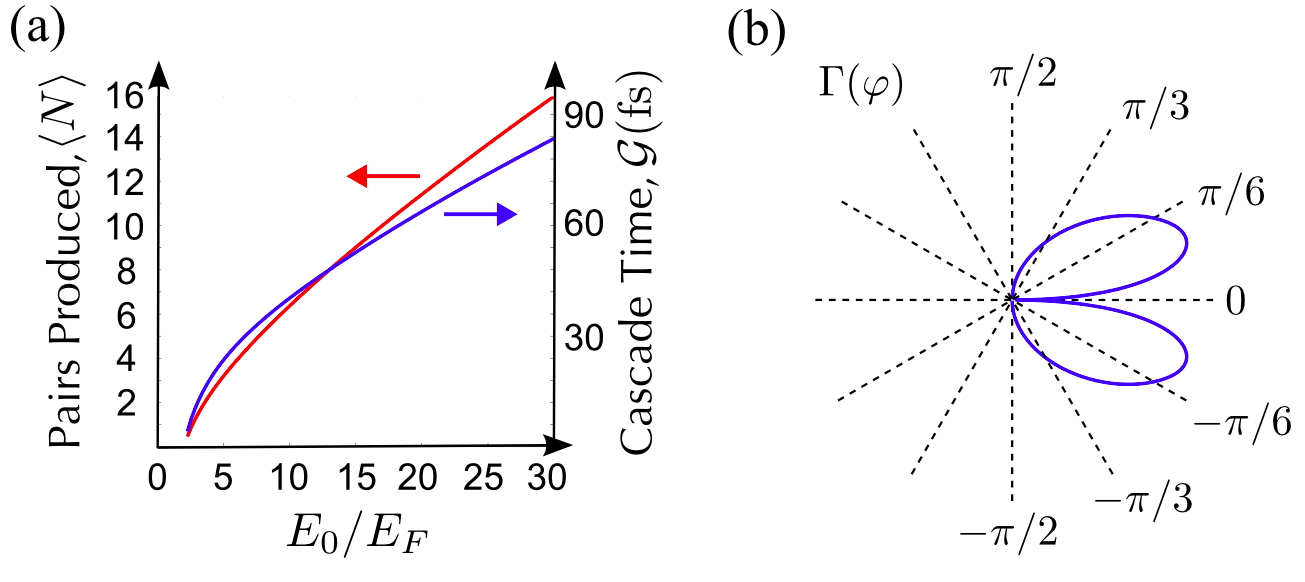}
\caption{ (a) Average net number of e-h pairs produced in the cascade triggered by a photo-excited electron with energy $E_0$ (red curve). Cascade duration, $\Delta t = \mathcal{G}/( \mu \, [{\rm eV}]){\rm fs}$, 
see Eq.(\ref{eq:deltat}) (blue curve).  (b) Angular dependence for the e-h excitation rate $\Gamma(\varphi)$,
where $\varphi$ is the angle between $\vec{k_1}$ and $\vec q$, see Fig.\ref{fig3}(a).}
\label{fig2}
\end{figure}

The dependence on excitation energy $E_0$ and Fermi energy $E_F$ provides clear experimental signatures of this relaxation mechanism.
In particular, the average number of e-h pairs produced in the cascade triggered by a {\it single photo-excited electron}, $\la N \ra$, is
\be\label{eq:avN}
\la N \ra = \int_{E_L}^{E_0} \frac{d \epsilon}{\la \Delta \epsilon \ra}, \quad \la \Delta \epsilon \ra= \frac{\mathcal{J}(\epsilon)}{\Gamma(\epsilon)},
\ee
where $\la \Delta \epsilon \ra$ is the average energy loss per step, $\mathcal{J}(\epsilon)$ and $\Gamma(\epsilon)$ are the IE energy-relaxation and scattering rates respectively (see Eq.(\ref{eq:Jnumerical})). Here $E_L \approx E_F$ is a low-energy cutoff corresponding to the energy below which IE processes are quenched; we use $E_L = 2 E_F$ (see discussion below). Fig.\ref{fig2}(a) indicates that $\la N \ra$ exceeds unity and grows quickly 
for $E_0$ above few $E_F$ (red curve).
Since $\la \Delta \epsilon \ra \sim E_F$, we find that $\la N \ra$ scales as $E_0/E_F$.  In particular, an approximately linear dependence $\la N \ra \approx 0.55 E_0/E_F$ is found for $E_0/E_F \gg 1$.


Similarly, the time it takes for the photo-excited electron to completely decay, $\Delta t$, 
also exhibits strong $E_0$ and $E_F$ dependence. This fairly short time, on the order of hundreds of femtoseconds, is
\be\label{eq:deltat}
\Delta t = \int_{E_L}^{E_0} \frac{d \epsilon}{ \mathcal{J}(\epsilon)} = \frac{ \mathcal{G} (E_0/E_F)}{ E_F [{\rm eV}]} \, {\rm fs}, 
\ee
where $\mathcal{G}$ is a dimensionless scaling function (blue curve in Fig. \ref{fig2}(a)). As shown in Fig. \ref{fig2}(a), $\mathcal{G}$ scales approximately linearly with $E_0/E_F$, yielding a $\Delta t$ that scales linearly with the excitation energy.
%
%
For a typical doping value of $E_F = 0.2\, {\rm eV}$ and initial photo-excited carrier energy $E_0 = 1\, {\rm eV}$  we find $\Delta t \approx  0.12 \, {\rm ps}$, far faster than typical electron-lattice cooling time scales found in graphene \cite{wong,macdonald, song12}. This separation of time scales means that the energy relaxation cascade occurs independently of electron-lattice cooling. 

Lastly, 
the angular distribution for impact excitation transitions is highly anisotropic. This produces a strong search-light-type structure peaked along
the preferred direction of momentum transfer shown in Fig. \ref{fig2} (b).


Our system is described by the Hamiltonian for $N=4$ species of massless Dirac particles,
\bea
&&
\mathcal{H}=  \sum_{\vec{k},i}  \psi^\dag_{\vec k,i} (  v\sigma\cdot\vec k ) \psi_{\vec k,i}  + \mathcal{H}_{\rm el-el}, 
\label{eq:hamiltonian}
\\
&&
\mathcal{H}_{\rm el-el} = \frac12\sum_{\vec q, \vec k,\vec{k'},i,j}V(\vec q) \psi^\dag_{\vec k + \vec q, i} \psi^\dag_{\vec{k'} - \vec q,j} \psi_{\vec {k'},j} \psi_{\vec k, i} 
.
\eea
%
Here $i,j=1...N$ and $V(\vec q)=2\pi e^2/|\vec q|\kappa$ is the Coulomb interaction. 
Importantly, transitions in a massless Dirac band governed by the Hamiltonian (\ref{eq:hamiltonian}) are subject to certain kinematical constraints \cite{aleiner,rana}. These constraints arise due to the combined effect of linear dispersion in two Dirac cones, $E_{\pm}(\vec p)=\pm v|\vec p|$, and momentum conserving character of carrier scattering. 
Here we analyze the simplest case of a two-body collision.
Each of the two particles participating in a collision can make transitions between states in the upper and lower Dirac cones which we denote by $+$ and $-$. Two kinds of transitions can be distinguished: intra-band transitions ($+\to +$ or $-\to -$) and inter-band transitions ($+\to -$ or $-\to +$). Since momentum change in any transition satisfies $||\vec p_1|-|\vec p_2||<|\vec p_1-\vec p_2|<|\vec p_1|+|\vec p_2|$, the intra-band transitions can only occur when the energy and momentum change are related by $|\Delta\epsilon|\le v|\Delta p|$, whereas the inter-band transitions are possible only when $|\Delta\epsilon|\ge v|\Delta p|$.

\begin{figure}
\includegraphics[scale=0.23]{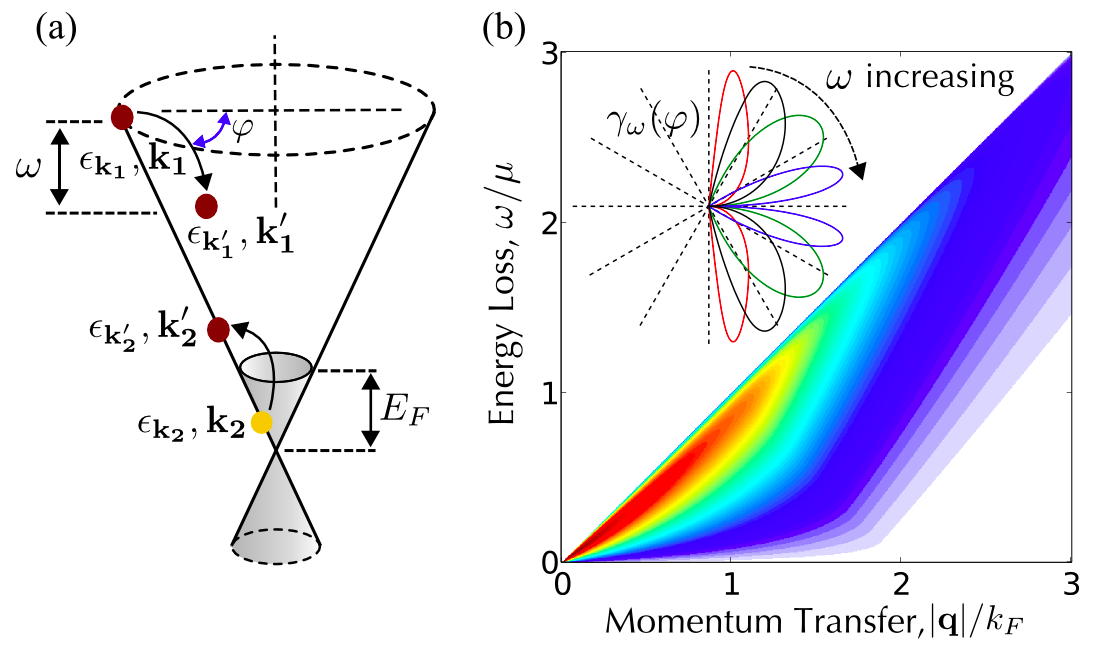}
\caption{(a) Kinematics of intraband carrier-carrier scattering in doped graphene. A photo-excited electron makes transition from $\vec{k_1}$ to $\vec{k_1'}$ by exciting an electron-hole pair from the Fermi sea from $\vec{k_2}$ to $\vec{k_2'}$. (b) 
Spectral function $R(\vec q, \omega)$ of particle-hole excitation 
as a function of momentum transfer and energy transfer per scattering event.
(inset) Angular distribution of normalized energy resolved transition rate (see Eq. \ref{eq:gammaomega}, for fixed values of $\omega/E_F = 0.2, 0.5, 1, 10$ (red, black, green, and blue curves respectively). The near-collinear character of scattering at high $\omega$ is manifested in narrowing of the angular distribution.}
\label{fig3}
\end{figure}

The scattering process of interest, pictured in Fig.\ref{fig3}(a), involves a photo-excited carrier 
with high energy and momentum $\epsilon_\vec{k_1}\gg E_F$, $|\vec k_1|\gg k_F$,  which is scattered to a lower energy state having momentum  $\vec k_1'$ with recoil momentum $\vec q = \vec{k_1} - \vec k_1'$ given to an  electron in the Fermi sea. The latter process results in a particle-hole pair excitation, as depicted by a transition from $\vec{k_2}$ to $\vec k_2'$ in Fig. \ref{fig3} (a). The transition rate for this process, evaluated by the standard Golden Rule  approach, takes the form
\bea\label{eq:W}
W_{\vec{k_1'},\vec{k_1}}  &=& \frac{2\pi N}{\hbar}\sum_{\vec q, \vec{k_2}, \vec{k_2'}} 
 f_\vec{k_2}( 1- f_{\vec{k_2'}}) F_{\vec{k_2}, \vec{k_2'}'}| \tilde V_\vec{q} |^2
\\\nonumber
&&\times \delta_{\vec{k_1'}, \vec{k_1} + \vec q} \delta_{\vec{k_2'}, \vec{k_2} - \vec q}  \delta (\epsilon_{\vec{k_1'}}-\epsilon_{\vec{k_1}}  + \epsilon_{\vec{k_2'}}-\epsilon_{\vec{k_2}})
.
 \eea
Here $f_{\vec k}$ is a Fermi function, and $ F_{\vec k, \vec k'}=|\la \vec k's'|\vec k s\ra|^2$ is the coherence factor ($s,s'=\pm$ label states in the electron and hole Dirac cones).  We treat the Coulomb interaction which mediates scattering between the photo-excited carrier and the carriers in the Fermi sea by accounting for dynamical screening in the RPA approximation:
\be \label{eq:RPA}
\tilde V_\vec{q} = \frac{V_{\vec q}^0}{\varepsilon(\omega,\vec q)}
,\quad
\varepsilon(\omega,\vec q)=1- V_\vec{q}^0 \Pi (\vec q, \omega)
,
\ee
where $V_{\vec q}^0 = 2\pi e^2/|\vec q|\kappa$ and $\varepsilon(\omega,\vec q)$ describes dynamical screening.
Here $\Pi$ is the polarization operator
\be
\Pi (\vec q, \omega) =N\sum_{\vec k, s, s'} F_{\vec{k}, \vec{k} + \vec{q}; ss' } \frac{ f(\epsilon_{\vec{k}, s}) - f(\epsilon_{{\vec{k}+\vec q}, s'})}{ \omega + \epsilon_{\vec{k}, s}- \epsilon_{\vec{k} + \vec{q}, s'}  + i0}
,
\ee
with the band indices $\{s, s'\} = \pm $. This includes both intra- ($s=s'$) and inter- ($s\neq s'$) band contributions \cite{dassarma}. 

For Eq.(\ref{eq:W}) to give a non-vanishing result, the transitions $\vec k_1\to \vec k_1'$, $\vec k_2\to \vec k_2'$ must occur in like pairs, both intra-band or both inter-band. Since the transition $\vec k_1\to \vec k_1'$  is restricted to be within a single band, the transition $\vec k_2\to \vec k_2'$ must also be intra-band. 
As a result, the relaxation of the photo-excited carrier via inter-band scattering is blocked, whereas intra-band scattering is allowed. Kinematical blocking of inter-band processes can in principle be relieved by three-body (or, higher-order) collisions which are not discussed here. Such processes may become important at high excitation power, however we expect the effect of such processes to be weak in the low excitation power regime discussed below.

As shown below, the typical energy of an excited pair is much smaller than the photo-excitation energy  $\epsilon_{\vec k_1}$. Anticipating this result, it is convenient to factorize the transition rate by expressing it through the spectrum of secondary pair excitations by following the standard procedure \cite{seperation}. We write $\delta(\epsilon_{\vec{k_1'}}-\epsilon_{\vec{k_1}}  + \epsilon_{\vec{k_2'}}-\epsilon_{\vec{k_2}}) = \int_{-\infty}^\infty d\omega \delta(\epsilon_{\vec{k_1'}}-\epsilon_{\vec{k_1}} +\omega) \delta(\epsilon_{\vec{k_2'}}-\epsilon_{\vec{k_2}}-\omega)$.
Next, we use the identity
\be
 f_\vec{k_2} (1- f_\vec{k_2'}) =  (  f_\vec{k_2} - f_\vec{k_2'}) \times \big( N(\epsilon_{\vec{k_2'}}-\epsilon_{\vec{k_2}}) + 1\big)
 ,
\ee
where $N(\omega) = 1/ (e^{\omega/k_BT} - 1)$ is the Bose function taken at the electron temperature.
Finally, we express the sum of $(  f_\vec{k_2} - f_\vec{k_2'}) \delta(\epsilon_{\vec{k_2'}} - \epsilon_{\vec{k_2}} - \omega)$ through a suitably defined susceptibility
\be\label{eq:chi''}
\chi''(\vec q, \omega) = N\sum_{\vec k} F_{\vec{k}, \vec{k} + \vec{q} } (  f_{\vec{k}} - f_{\vec{k}+\vec q}) 
\delta(\epsilon_{\vec{k+q}}-\epsilon_{\vec{k}} - \omega)
,
\ee
%
which can also be written as $\chi''(\vec q, \omega)= - \frac1{\pi}\Im\Pi(\vec q,\omega)$.
This yields a compact and intuitive expression for the total scattering rate: 
\bea\label{eq:Gamma_total}
\Gamma =\sum_{\vec{k_1'}} W_{\vec k_1', \vec k_1} ( 1- f_{\vec k_1'})  F_{\vec{k_1}, \vec k_1' } =\int_{-\infty}^{\infty}\!\!\! d\omega  P(\omega)
,
\\ 
P (\omega) =
%
A \sum_{\vec q} | \tilde V_{\vec{q}} |^2 F_{\vec{k_1}, \vec k_1' }\chi''(\vec q, \omega) 
\delta(\epsilon_{\vec k_1'}-\epsilon_{\vec k_1} +\omega)
,
\label{eq:p}
\eea
where $A=\frac{2\pi}{\hbar}[N(\omega)+1)][ 1 - f(\epsilon_{\vec k} - \omega)]$ and $\vec k_1'=\vec k_1-\vec q$.

As we show below, the typical energy and momentum transferred per scattering is of order $E_F$ and $E_F/v$, respectively. These values are much smaller than those of the photo-excited electron. We can therefore approximate $F_{\vec k_1, \vec k_1' }\approx 1$, $f(\epsilon_k - \omega) \approx 0$ and write the delta function as
\be
\delta(\epsilon_{\vec k_1'}-\epsilon_{\vec k_1} +\omega)\approx \delta (v|\vec q| \cos \varphi - \omega) 
,
\ee
where $\varphi$ is the angle between $\vec k_1$ and $\vec q$. The approximation $|\vec q|\ll |\vec k_1|$, $|\omega|\ll v|\vec k_1|$ is appropriate under realistic conditions:
for example, visible light frequencies translate to $\epsilon_{\vec k}=hf/2 = 750 \, {\rm meV}$ which is considerably larger than $E_F$ for typical doping values. Eq.(\ref{eq:Gamma_total}) then yields the angle dependent transition rate
\bea
&&\Gamma(\varphi)=\int_{-\infty}^{\infty}\!\!\! d\omega \gamma_\omega (\varphi)
,
\label{eq:gammaphi}
\\
&&
\gamma_\omega (\varphi) = \int \frac{d^2q}{(2\pi)^2} R(\vec q,\omega)\delta (v|\vec q| \cos \varphi - \omega)
, 
\label{eq:gammaomega}
\eea
where $R(\vec q,\omega)=A | \tilde V_{\vec{q}} |^2 \chi''(\vec q, \omega) $. 
We evaluate the spectral function $R(\vec q, \omega)$ using the RPA-screened interaction, Eq.(\ref{eq:RPA}),and susceptibility expressed through the polarization function 
from Ref. \cite{dassarma} and the interaction parameter $\alpha = e^2/(\kappa \hbar v) = 0.73$. 
The angular distribution, $\Gamma (\varphi)$, as well as the energy resolved distribution, $\gamma_\omega (\varphi)$,
feature interesting angular patterns (see Fig.\ref{fig2} (b) and Fig.\ref{fig3} inset). Note in particular a sharp search-light-type structure corresponding the preferred direction of momentum transfer, $\vec q$, in the IE process. The peaks move closer to $\varphi=0$ as $\omega$ increases, indicating that the carrier-carrier scattering with a high energy transfer is nearly collinear. This is analogous to radiation pattern for an ultra-relativistic particle becoming focused along particle velocity\cite{jackson}.

The angular patterns indicate that IE does not scatter into the angles $|\varphi |\geq \pi/2$ and vanishes at $\varphi=0$. The origin of these features can be understood from the spectral function $R(\vec q, \omega)$ shown in Fig. \ref{fig3} (b) which peaks close to $\omega = v|\vec q|$ but vanishes for both $\varphi = \arccos (\omega/ v|\vec q|) = 0$ and $\varphi = \arccos (\omega/ v|\vec q|) =  \pm \pi/2$. 
Scattering angles $|\varphi |\geq \pi/2$ correspond to negative $\omega$, i.e. to the IE processes in which the photo-excited carrier gains energy from the Fermi sea. These processes are suppressed by the Bose factor in $A$. Na\"ively one would expect that $\varphi=0$, which corresponds to perfectly collinear scattering, would maximize transition rates since the phase space for particle-hole excitations in $\chi''$ diverges when $v|\vec q| = \omega$. However, because the same divergence also appears in $\Pi$, the RPA-screened interaction makes $R(\vec q, \omega)$ vanish on the cone $\omega=v|\vec q|$. We note that this suppression of perfectly collinear scattering is highly sensitive to the approximation employed to describe dynamically screened interaction. The angular patterns $\Gamma(\varphi)$ and $\gamma_\omega(\varphi)$ can therefore be used as a test of the validity of the RPA approximation.

The same approach can be used to obtain the energy spectrum of pair excitations. In the following, however, we study the {\it full energy dependence} of $P(\Delta \epsilon)$ not limiting ourselves to the asymptotic behavior at high photo-excited energies.
Using a Jacobian to convert the delta function in energy to a delta function in angles in Eq.(\ref{eq:p}), we perform the angular integral in Eq.(\ref{eq:p}) to obtain
%
\be\label{eq:pav}
P(\Delta \epsilon) = \int_0^{\infty}  \frac{2 |\vec{k_1}| - (\Delta \epsilon /v\hbar) - |\vec q| \cos \varphi}{ (2 \pi)^2|\vec{k_1}| |\vec q| \hbar  v \sin \varphi}  R(\vec q, \Delta \epsilon) q dq,
\ee
where $\varphi$ is the angle between $\vec{k_1}$ and $\vec q$ and satisfies $( |\vec{k_1}|^2 - 2  |\vec{k_1}| |\vec{q}|\cos\varphi + |\vec{q}|^2)^{1/2} -  |\vec{k_1}| = \Delta \epsilon/(v\hbar)$.
Numerically integrating Eq.(\ref{eq:pav}) and taking $\epsilon \gg k_BT$ yields transition probabilities
$P(\Delta \epsilon)$ shown in Fig. \ref{fig1}(c) for different initial photo-excited energies $\epsilon_i = \epsilon_{\vec k_1}$. We find that $P(\Delta \epsilon)$ peaks close to $\Delta \epsilon \approx E_F$ and decays rapidly for $\Delta \epsilon \gg E_F$. This non-monotonic dependence arises from the competition between the available phase space, which grows with $\Delta \epsilon$, and the Coulomb Interaction form factor, which decreases with $|\vec q|$.

The efficiency of IE scattering noted above can be linked to the large values of $E_F$ in graphene. 
The relation between efficiency and $E_F$ can be clarified by simple dimensional analysis.
%
We note that 
$P(\Delta \epsilon)$ depends on $\Delta \epsilon$ essentially via the dimensionless parameter $x=\Delta \epsilon/E_F$. This is clearly seen e.g. from pair excitation spectrum shown
 for different values of initial energy $\epsilon_i = \epsilon_{\vec k_1}$ in Fig.\ref{fig1}(c): the width and profile of $P(\Delta\epsilon)$ has a very weak dependence on $\epsilon_i$.
This can be captured by writing
the scattering rate $\Gamma$ (Eq.(\ref{eq:Gamma_total})) as well as energy relaxation rate $\mathcal{J}=\int_{-\infty}^\infty \Delta \epsilon P(\Delta \epsilon) d\Delta \epsilon$ 
in the form
%
%
%
%
\be
\Gamma(\epsilon) = \frac{E_F}{\hbar} \int_0^{\epsilon/E_F} \hspace{-7mm}\tilde{P}(x) dx, \quad \mathcal{J}(\epsilon) = \frac{E_F^2}{\hbar} \int_0^{\epsilon/E_F} \hspace{-7mm}x\tilde{P}(x) dx,
\label{eq:Jnumerical}
\ee
where we introduced dimensionless $\tilde P(x) = \hbar P(\Delta \epsilon)$. 
Using  $P(\Delta \epsilon)$ evaluated from Eq.(\ref{eq:pav}), we obtain the fast impact excitation energy relaxation rate shown in Fig. \ref{fig1}(b) for $E_F = 0.2 \, {\rm eV}$. Comparing $\mathcal{J}(\epsilon)$ from impact excitation with the energy relaxation rate arising from the emission of optical phonons, $\mathcal{J}_{\rm el-ph}$ \cite{appendix}, we obtain a branching ratio $\mathcal{J}/\mathcal{J}_{\rm el-ph}$ shown in the inset if Fig. \ref{fig1}(b). While $\mathcal{J}_{\rm el-ph}$ does not depend on carrier density \cite{appendix}, $\mathcal{J}$ does. As a result, gate voltage 
can be used to tune the branching ratio $\mathcal{J}/\mathcal{J}_{\rm el-ph}$ by up to an order of magnitude as illustrated in the inset of Fig. \ref{fig1}(b).


The $E_F$ dependences in Eq.(\ref{eq:Jnumerical}) manifests in observables such as
the average number of secondary e-h pairs produced in a single photo-excitation
cascade, $\la N \ra$ and its total cascade time, $\Delta t$. 
These quantities are related via $\la N \ra = \int_0^{\Delta t} \Gamma dt$. Using $d\epsilon/ dt = -\mathcal{J} (\epsilon)$ combined with Eq.(\ref{eq:Jnumerical}) we obtain Eq.(\ref{eq:avN}) for $\la N \ra$ and Eq.(\ref{eq:deltat}) for the cascade time. In both cases, we have used a low energy cutoff for the energy below which IE processes are quenched, $E_L \approx E_F$. Below the energy $E_L$, the relaxation and scattering of the carrier from impact excitation slows dramatically and other relaxation processes dominate, for example energy relaxation via the emission of acoustic phonons. In evaluating Eqs.(\ref{eq:avN}),(\ref{eq:deltat}), we used the value $E_L=2 E_F$ 
below which the predicted value $\Delta t$
rapidly increases. The scaling 
 of $\Delta t$ and $\la N \ra$ with both excitation energy $E_0$ and doping in Fig. \ref{fig2}(a) are clear experimental signatures of IE. 
Currently, particle-hole pair production and cascade times are the subject of intense experimental interest \cite{beard00,schaller04,klaas}.

In summary, the impact excitation 
scattering is predicted to be a dominant carrier relaxation pathway in doped graphene. Multiple pair excitations are produced in a cascade triggered by a single photon, featuring an approximately linear scaling of the number of generated pairs and total cascade time with photo-excitation energy.
These dependences, as well as a sharply peaked angular distribution of e-h pairs, provide clear experimental signatures for IE-dominated cascade.  
Strong gate dependence of the
cascade parameters affords 
a useful knob for the control of ultrafast scattering processes in graphene.

We acknowledge financial support from the NSS program, Singapore (JS), the Office of Naval Research Grant No. N00014-09-1-0724 (LL), and Fundacio Cellex Barcelona (KJT, FK).

\section{Appendix A: Energy relaxation from the emission of optical phonons }

An alternative channel for energy relaxation of photo-excited carriers occurs through the emission of optical phonons and gives an energy relaxation rate of  $\mathcal{J}_{\rm el-ph}$. The transition rate of this process \cite{tse08} can be described by Fermi's golden rule
\bea
W_{\vec{k}',\vec{k}}^{\rm el-ph}  &=& \frac{2\pi N}{\hbar}\sum_{\vec q} |M(\vec{k}',\vec{k})|^2  \\ \nonumber
&& \delta \big(\Delta\epsilon_{\vec{k}', \vec{k}}  + \omega_\vec{q}  \big) \delta_{\vec{k}', \vec{k} + \vec q} (N({\omega_\vec{q}}) + 1),
\eea
where $\Delta\epsilon_{\vec{k}', \vec{k}} = \epsilon_{\vec k'} - \epsilon_{\vec k}$, $\omega_\vec{q}=\omega_0 = 200\, {\rm meV}$ is the optical phonon dispersion relation, and $N({\omega_{\vec q}})$ is a Bose function. Here $\vec k$ is the initial momentum of the photo-excited electron, $\vec k'$ is the momentum it gets scattered into, and $\vec q$ is the momentum of the optical phonon.The electron-phonon matrix element $M(\vec{k}',\vec{k})$ is
\be
|M(\vec{k}',\vec{k})|^2 = g_0^2 F_{\vec{k}, \vec{k}' }, \quad g_0 = \frac{2\hbar^2v}{\sqrt{2\rho \omega_0 a^4}},
\ee
where $ F_{\vec{k}, \vec{k}' }$ is the coherence factor for graphene, $g_0$ is the electron-optical phonon coupling constant \cite{macdonald}, $\rho$ is graphene's mass density, and $a$ is the distance between nearest neighbor carbon atoms. The energy-loss rate of the photo-excited carrier at energy $\epsilon$ due to the emission of an optical phonon is
\be
\mathcal{J}_{\rm el-ph}(\epsilon) = \sum_{\vec{k'}} W_{\vec{k}', \vec{k}}^{\rm el-ph} (\epsilon_k' - \epsilon)\big[ 1- f(\epsilon_\vec{k'})\big]  .
\ee
Integrating over $\vec q$ and $\vec k'$ we obtain  
\be
\mathcal{J}_{\rm el-ph}(\epsilon) = \frac{\pi N}{\hbar} \omega_0 g_0^2 \big[ 1- f(\epsilon - \omega_0)\big]  (N(\omega_0)+1)\nu(\epsilon - \omega_0), 
\ee
where $\nu(\epsilon)=\epsilon/(2\pi v^2\hbar^2)$ is the electron density of states in graphene. . Hence, $\mathcal{J}_{\rm el-ph}(\epsilon)$ varies linearly with the photo-excited carrier energy $\epsilon > \omega_0$ and vanishes for $\epsilon<\omega_0$. Because the electron-phonon coupling with optical phonon is a constant, this result is to be expected from the increased phase space to scatter into at higher photo-excited carrier energy.

To get an estimate of the energy relaxation rate, we estimate $(N(\omega_0)+1) \approx 1$ and $1- f(\epsilon - \omega_0) \approx \Theta(\epsilon - \omega_0- E_F)$ and use $\rho = 7.6 \times 10^{-11} \, {\rm kg} \, {\rm cm}^{-2}$ and $a = 1.42 \AA$ to obtain
\be
\mathcal{J}_{\rm el -ph} (\epsilon) 
\approx \frac{\epsilon - \omega_0}{734 \, {\rm fs}} , \quad  \epsilon> E_F + \omega_0.
\label{eq:Jph}
\ee
Using this we plot the energy relaxation rate for photo-excited electrons from the emission of optical phonons in Fig .\ref{fig1} (b). As shown in Fig. \ref{fig1} (b), this channel is far smaller than the impact excitation channel described in the main text for typical dopings and photo-excited carrier energies.


\begin{thebibliography}{99}
\bibitem{kampfrath05} T. Kampfrath, L. Perfetti, F. Schapper, C. Frischkorn, and M. Wolf, Phys. Rev. Lett. {\bf 95}, 187403 (2005).
\bibitem{butscher07} S. Butscher, F. Milde, M. Hirtschulz, E. Mali\'c, and A. Knorr, App. Phys. Lett. {\rm 91}, 203103 (2007). 
\bibitem{george08} P. A. George, J. Strait, J. Dawlaty, S. Shivaraman, M. Chandrashekhar, F. Rana, and M. G. Spencer, NanoLett. {\bf 8}, 4248 (2008).
\bibitem{vasko08} F. T. Vasko, and V. Ryzhii, Phys. Rev. B {\bf 77}, 195433 (2008). 

\bibitem{breusing09} M. Breusing, C. Ropers, and T. Elsaesser, Phys. Rev. Lett. {\bf 102}, 086809 (2009).



\bibitem{lui10} C. H. Lui, K. F. Mak, J. Shan, and T. F. Heinz, Phys. Rev. Lett., {\bf 105} 127404 (2010).


\bibitem{winzer10} T. Winzer, A. Knorr, and E. Mali\'c, NanoLett., {\bf 10} 4839 (2010).

\bibitem{kim11} R. Kim, V. Perebeinos, P. Avouris, Phys. Rev. B, {\bf 84} 075449 (2011).
\bibitem{winnerl11} S. Winnerl, M. Orlita, P. Plochacka, P. Kossacki,  M. Ptemski, T. Winzer, E. Malic, A. Knorr, M. Sprinkle, C. Berger, W. A. de Heer, H. Schneider, and M. Helm,
Phys. Rev. Lett., {\bf 107} 237401 (2011).
\bibitem{winzer12} T. Winzer, and E. Mali\'c, Phys. Rev. B, {\bf 85} 241404 (2012).


\bibitem{aleiner} M. S. Foster, and I. L. Aleiner, Phys. Rev. B, {\bf 79} 085415 (2009).



\bibitem{rana} F. Rana, Phys. Rev. B, {\bf 76} 155431 (2007).
 
 

\bibitem{Guinea1994} J. Gonzalez, F. Guinea, M. A. H. Vozmediano, Nuclear Physics {\bf B424}, 595 (1994).
\bibitem{Geim2011} D. C. Elias, R. V. Gorbachev, A. S. Mayorov, S. V. Morozov, A. A. Zhukov, P. Blake, L. A. Ponomarenko, I. V. Grigorieva, K. S. Novoselov, F. Guinea, and A. K. Geim, Nature Physics {\bf 7},  701 (2011).
\bibitem{Lanzara2006} 
S. Y. Zhou, G.-H. Gweon, J. Graf, A. V. Fedorov, C. D. Spataru, R.D. Diehl, Y. Kopelevich, D.-H. Lee, Steven G. Louie, A. Lanzara, Nature Physics {\bf 2}, 595 (2006).
\bibitem{Rotenberg2007} A. Bostwick, T. Ohta, T. Seyller, K. Horn, E. Rotenberg, Quasiparticle dynamics in graphene. Nat. Phys. {\bf 3}, 36 (2007).
\bibitem{polini_privatecommunication} M. Polini, private communication


\bibitem{appendix} See Appendix.
\bibitem{tse08} W.-K. Tse, E. H. Hwang, and S. Das Sarma, App. Phys. Lett., {\bf 93}, 023128 (2008).
 

 
 
 
 \bibitem{macdonald} R. Bistritzer, and A. H.  MacDonald, Phys. Rev. Lett., {\bf 102}, 206410 (2009). 

\bibitem{wong} W.-K. Tse, and S. Das Sarma, Phys. Rev. B, {\bf 79}, 235406 (2009). 

\bibitem{song12} J. C. W. Song, M. Y. Reizer, and L. S. Levitov, Phys. Rev. Lett., {\bf 109} 106602 (2012).
 
 \bibitem{seperation} G. Giuliani, and J. J. Quinn, Phys. Rev. B, {\bf 26} 4421 (1982).
 

\bibitem{dassarma} E. H. Hwang, and S. Das Sarma, Phys. Rev. B, {\bf 75} 205418 (2007).


\bibitem{jackson} J. D. Jackson, {\it Classical Electromagnetism},  pp. 668-669, Third Edition, Wiley (1999). 

 \bibitem{schaller04} R. D. Schaller, and V. I. Klimov, Phys. Rev. Lett., {\bf 92} 186601 (2004).
\bibitem{beard00} M. C. Beard, G. M. Turner, and C. A. Schmuttenmaer, Phys. Rev. B, {\bf 62} 15764 (2000).
\bibitem{klaas} K.J. Tielrooij, J.C.W. Song, S.A. Jensen, 5 A. Centeno, A. Pesquera,
A. Zurutuza Elorza, M. Bonn, L.S. Levitov, and F.H.L. Koppens, to be published





\end{thebibliography}
\end{document}